\documentclass[final]{ias2}

\usepackage{graphicx} 
\usepackage{multirow}
\usepackage{array} 

\usepackage{hyperref} 

\begin{document}

\markboth{Thermal conductivity of nonlinear waves in disordered chains}{Sergej Flach, et. al.}

\title{Thermal conductivity of nonlinear waves in disordered chains}

\author[sin]{Sergej Flach} 
\email{flach@pks.mpg.de}
\author[sin,ain]{Mikhail Ivanchenko} 
\email{ivanchen@pks.mpg.de}
\author[sin]{Nianbei Li}
\email{nianbei@pks.mpg.de}
\address[sin]{Max-Planck-Institut f\"ur Physik komplexer Systeme, N\"othnitzer Str. 38, 01187 Dresden, Germany}
\address[ain]{Theory of Oscillations Department, University of Nizhniy Novgorod, Russia}

\begin{abstract}
We present computational data on the thermal conductivity of nonlinear waves in disordered chains.
Disorder induces Anderson localization for linear waves and results in a vanishing conductivity.
Cubic nonlinearity restores normal conductivity, but with
a strongly temperature-dependent conductivity $\kappa(T)$. We find indications for an asymptotic low-temperature
$\kappa \sim T^4$ and intermediate temperature $\kappa \sim T^2$ laws. These findings 
are in accord with theoretical studies of wave packet spreading, 
where a regime of strong chaos is found to be intermediate, followed by
an asymptotic regime of weak chaos (EPL 91 (2010) 30001).
\end{abstract}


 
\maketitle


\section{Introduction}

In the absence of nonlinearity (or many-body interactions in quantum systems) all eigenstates - normal modes (NM) -  in one-dimensional random lattices with disorder are spatially localized. This is Anderson localization \cite{PWA58}, which has been discovered fifty years ago in disordered crystals as a localization of electronic wavefunctions. It can be interpreted as an interference effect between multiple scattering events of the electron by random defects of the potential. Recent experiments on the observation of Anderson localization were performed with light propagation in spatially random optical media \cite{Exp, Exp2}, with noninteracting Bose-Einstein condensates expanding in random optical potentials \cite{Billy,Roati},   and with wave localization in a microwave cavity filled with randomly distributed scatterers \cite{microwave}.

In many situations nonlinear terms in the wave equations (respectively, many body interaction terms in quantum systems) have to be included. Thus, a fundamental question which has attracted the attention of many researchers is: what happens to an initial excitation of arbitrary shape in a nonlinear disordered lattice. 
Nonlinearity renormalizes excitation frequencies, and induces interactions between NMs. Numerical studies show that wave packets spread subdiffusively and Anderson localization is destroyed \cite{PS08,fks09,skfk09,lbksf10}. In the regime of strong nonlinearity, far from where it can be treated perturbatively, new localization effects of selftrapping occur \cite{KKFA08}. A theoretical explanation of the subdiffusive spreading was offered in Refs. \cite{fks09,skfk09,F10}.  It is based on the fact that the considered models are in general nonintegrable. Therefore deterministic chaos will lead to an incoherent spreading. Estimates of the excitation transfer rate across the packet tail are obtained by calculating  probabilities of mode-mode resonances inside the packet. Some predictions of this approach include the effect of different degrees of nonlinearity and were successfully tested in \cite{SF10}. 
First experimental data were recently presented for repulsively interacting Bose-Einstein condensates in quasiperiodic potentials \cite{deissler,lucioni}
which, among others, confirm the basic theoretical findings. 

While the above wave packets are spreading into an 'empty' system, another set of related problems concerns the thermal conductivity at finite temperature $T$ (i.e. at finite
energy density $\varepsilon \sim T$) \cite{slrlap03,ad08}. In the absence of nonlinearity Anderson localization of NMs leads to a vanishing
conductivity. It was observed that even a small amount of anharmonicity leads to a diffusive transport of energy, and a finite thermal conductivity $\kappa$ \cite{adjll08}.
Some numerical studies indicated that for low temperatures $\kappa \sim T^{1/2}$ \cite{adjll08} which would indicate a singularity at zero temperatures. Other expectations claim
that heat conductivity vanishes strictly for weak enough (but still finite) anharmonicity \cite{ad08}. 

In this report we use the theoretical results on wave packet spreading and derive a connection between $\kappa$ and $T$. We predict that for low temperatures $\kappa\sim T^4$ which
corresponds to the regime of weak chaos. At higher temperatures an optional intermediate regime of strong chaos may lead to $\kappa \sim T^2$.
We perform numerical simulations which indicate that our predictions are plausible.

\section{Models}

We study two different Hamiltonian models. The first model describes interacting anharmonic oscillators in a quartic Klein-Gordon (KG) chain, given as
\begin{equation}
\mathcal{H}_K = \sum_l \frac{p_l^2}{2}+\frac{{\tilde \epsilon}_l}{2} u_l^2 + \frac{1}{4}u_l^4 + \frac{1}{2W}(u_{l+1}-u_l)^2 \label{eq:KG}
\end{equation}
where $u_l$ and $p_l$ are respectively the generalized coordinate/momentum on the site $l$,
and $\tilde{\epsilon_l}$ are the disordered potential strengths chosen uniformly in 
the interval $[1/2,3/2]$. 
Likewise, $\partial_t^2 u_l = - \partial \mathcal{H}_K / \partial u_l$ generates the equations of motion 
\begin{equation}
\ddot{u}_l = -{\tilde \epsilon}_l u_l - u_l^3 + \frac{1}{W}(u_{l+1}+u_{l-1} - 2u_l) \;. \label{eq:KGEOM}
\end{equation}

By neglecting the nonlinear terms, the KG model (\ref{eq:KG}) reduces to a linear eigenvalue problem. 
This leads to a set of NM amplitudes, $A_{\nu,l}$, with NM squared frequencies $\omega_{\nu}^2 \in [1/2, 3/2 + 4/W]$ and  
$\Delta=1+4/W$ being the width of the squared eigenfrequency 
spectrum.
The NM asymptotic spatial decay is given by $A_{\nu, l} \sim e^{-l/\xi(\omega^2_\nu)}$ where $\xi(\omega^2_\nu) $ is the localization length. 
It is approximated \cite{KM93} in the limit of weak disorder ($W \ll 1$) as $\xi{(\omega^2_\nu)} \le  96 W^{-2}$. 
The NM participation number $P_{\nu} = 1/\sum_{l}A^4_{\nu,l}$ characterizes the NM spatial extent. 
An average measure of this extent is the localization volume $V$, which is on the order of $3.3 \xi(0)$ for weak disorder
and unity in the limit of strong disorder \cite{KF10}. The average squared frequency spacing of NMs within a localization volume is then $d \approx \Delta/V$. The two squared frequency 
scales $d$ and $\Delta$  with $ d < \Delta$ are thus expected to determine the dynamics details in the presence of nonlinearity. 

Nonlinearity induces an interaction between NMs. Since all NMs are exponentially localized in space, each of them is effectively coupled to a finite number of neighbor 
modes, \textit{i.e.} the interaction range is finite. However, the strength of this coupling is proportional to a characteristic energy density,
$\varepsilon$. The squared frequency shift due to the nonlinearity is then $\delta \sim \varepsilon$.

The KG model is suitable for studies of the dynamics for weak and intermediate disorder $ W < 20$, since otherwise the coupling strength between nearest neighbor
oscillators becomes too weak and required computation times too long. In the limit of strong disorder the NMs become localized on a single site, and will be coupled
by nonlinearity to their nearest neighbors. Therefore this limit can be modeled by a chain of harmonic oscillators with random frequencies and purely anharmonic
nearest neighbor coupling:
\begin{equation}
\mathcal{H}_{FSW} = \sum_l \frac{p_l^2}{2}+\frac{{\tilde \epsilon}_l}{2} u_l^2 + \frac{1}{4}(u_{l+1}-u_{l})^4  \label{eq:FSW}
\end{equation}
which is an example from a class of models studied by Fr\"ohlich, Spencer and Wayne \cite{fsw86} and therefore called here the FSW model.
It follows qualitatively from the KG model by first transforming into NM space, dropping secular terms \cite{F10},
and finally through the limits of strong disorder, and of high temperatures (energies). Its quartic potential part however is invariant
under homogeneous coordinate shifts, contrary to the KG case.

\section{Thermal conductivity}

The thermal conductivity is measured by a standard approach, when the thermal baths attached to the ends generate a temperature gradient and heat current  along the chain \cite{slrlap03}.
For the KG model we implement Langevin thermostats
with two
next-to-end atoms  coupled to Langevin heat
baths with
$\ddot{u}_{1,N}=-\partial{H}/\partial{u_{1,N}}+\xi_{1,N}-\lambda\dot{u}_{1,N}$ with
white noise $\left<\xi_{1,N}\right>=0$ and
$\left<\xi_{1,N}(t)\xi_{1,N}(0)\right>=2\lambda T_{1,N}\delta(t)$ where
$\left<\cdot\right>$ denotes ensemble averages. $T$ is the
temperature, and $\lambda$ denotes the
coupling strength between system and heat bath.
For the FSW model we use
Nos\'e-Hoover thermostats adding the terms $-\zeta_\pm\dot{u}_{1,N}$ to the respective equations of motion, where $\dot{\zeta}_\pm=\dot{u}_{1,N}^2/T_{1,N}-1$. 
\begin{figure}[ht]
\begin{center}
\includegraphics[width=0.9\columnwidth]{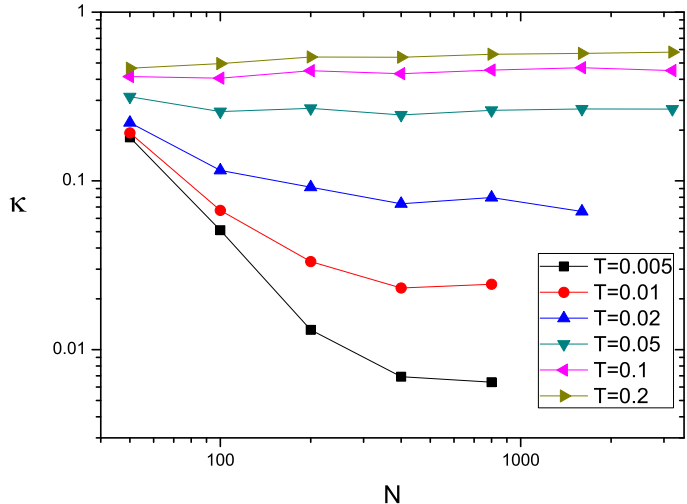}
\caption{KG chain: the size-dependent thermal conductivity $\kappa_N$ versus $N$ for different
temperatures and $W=2$.}
\label{fig2}
\end{center}
\end{figure}

For the KG model the heat flux along the chain is defined as the 
time average of $j_{KG}=-\frac{1}{2W}\sum\limits_l(\dot{u}_{l+1}+\dot{u}_l)(u_{l+1}-u_{l})$
\cite{slrlap03}. 
For the FSW model the heat flux follows from the time average of 
$j_{FSW}=-\frac{1}{2}\sum\limits_l(\dot{u}_{l+1}+\dot{u}_l)(u_{l+1}-u_{l})^3$ \cite{slrlap03}.
The thermal conductivity coefficient reads $\kappa=j N/(T_N-T_1)$ then \cite{slrlap03}. We make use of the mean temperature $T=(T_N+T_1)/2$  as a parameter corresponding to the energy density 
$\langle \varepsilon \rangle=T$,  and $(T_N-T_1)/T=0.5$ further on.

\section{Predictions}

The numerical evolution of a wave packet spreading into an empty system \cite{PS08,fks09,skfk09,lbksf10} yields subdiffusive growth of the second moment.
This implies that the diffusion rate is depending on the energy density (temperature). 
A theoretical approach \cite{fks09,skfk09,F10} postulates that the wave packet evolves chaotically, and estimates the diffusion rate for the KG model 
to be \cite{F10}
\begin{equation}
D \sim  \varepsilon^2 \mathcal{P}^2(\varepsilon)
\end{equation}
with the resonance probability \cite{KF10} 
\begin{equation}
\mathcal{P}(\varepsilon) \approx 1-{\rm e}^{-a \varepsilon/d}
\label{approxpp}
\end{equation}
and $a$ being an unknown constant of order one.

For small temperatures $\varepsilon \sim T$. Therefore, and assuming that the thermal conductivity is proportional to the diffusion rate \cite{lbksf10}, we conclude
that for the KG model
\begin{equation}
 \kappa \sim T^2 \left( 1-{\rm e}^{-b T /d} \right)^2
\end{equation}
with $b$ another constant of order one.
In particular we find that 
\begin{eqnarray}
\kappa \sim T^4 \;,\; T \ll d \;{\rm (weak\; chaos)} \\
\kappa \sim T^2 \;,\; T \gg d \;{\rm (strong\; chaos)}
\end{eqnarray}
Note that for larger energy densities (temperatures) the KG model enters the regime of selftrapping, and the above predictions do not apply
anymore. The selftrapping regime is expected when $\varepsilon\sim T \approx 2/(3W) $ \cite{lbksf10}. It
is the asymptotic high temperature limit of the KG model, where the thermal conductivity should drop with increasing temperature \cite{ad08}.
Thus, depending on the strength of disorder, we expect that the thermal conductivity grows 
with temperature as $T^4$, then (for not too strong disorder) it grows
as $T^2$, and finally goes through a maximum and starts to decrease with further increase of temperature.
The three phases are summarized in the inset of Fig.3 in Ref. \cite{lbksf10}.

The application of the above approach to the FSW model leads to $d\sim 1$ and to the absence of a selftrapped regime.
Instead, for large temperatures, the leading order anharmonic potential terms of the FSW model bring it close to the momentum conserving
$\beta$-Fermi-Pasta-Ulam chain (FPU). The FPU chain has a thermal conductivity which diverges with increasing system size \cite{slrlap03}.
Therefore $\kappa$ will increase with system size also for the FSW model for large temperatures, until the harmonic oscillator potential parts
can not be neglected anymore, leading to a final saturation of the thermal conductivity with further system size increase. 
Indeed, the harmonic oscillator potential parts are violating total momentum conservation, and therefore the thermal conductivity
stays finite for large enough system size \cite{lbksf10,ad08}. Since the momentum conserving terms of the FSW model become more dominant at larger temperatures,
it follows that
the higher the temperature, the larger the system sizes are needed, and the larger the saturated conductivity value. The crossover
to this pseudo-FPU regime can be expected when the temperature is of order unity (since all relevant FSW model parameters are of order unity).
Therefore we conclude, that we will not see a pronounced regime of strong chaos, instead the regime of weak chaos at low temperatures should
gradually transform into the pseudo-FPU regime.

\section{Numerical results}

The thermal conductivity is computed along the above lines, with additional averaging
over disorder realizations.

\subsection{KG chain}

\begin{figure}[ht]
\begin{center}
\includegraphics[width=0.9\columnwidth]{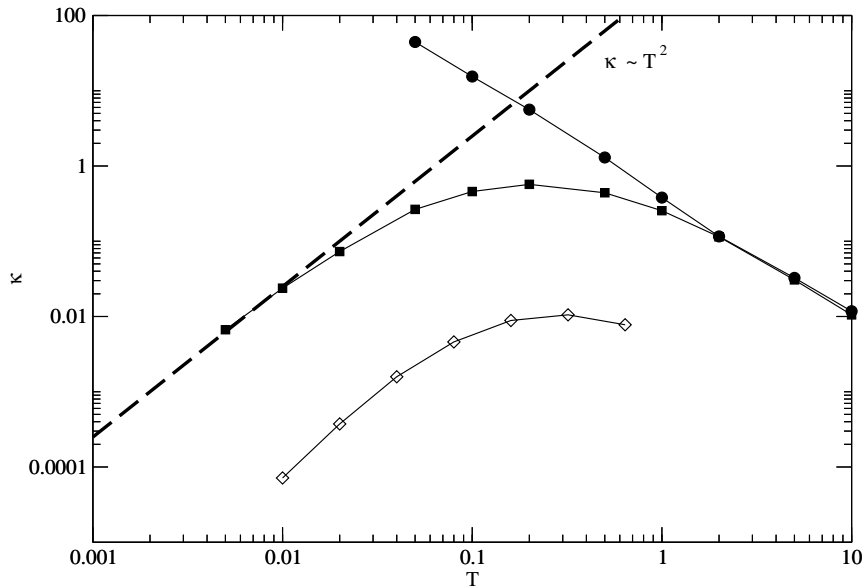}
\caption{KG chain: $\kappa(T)$ for $W=2$ (filled squares). For comparison we also show the
data for $\tilde{\epsilon}_l\equiv 1$ (filled circles). 
Thin solid lines guide the eye.
The dashed line corresponds to the power law
$T^2$. 
The stronger disorder case $W=6$ corresponds to the open diamond data points.}
\label{fig3}
\end{center}
\end{figure}

We first present results for the disorder strength $W=2$. In that case the localization
length $\xi \approx 25$ and the localization volume $V \approx 75$. The thermal conductivity
$\kappa_N$ is plotted for various temperatures versus system size $N$ in Fig.\ref{fig2}.
For large temperatures $T \geq 0.1$ the conductivity reaches its saturated level $\kappa$ at system
size $N \approx \xi$. 
For the linear wave equation, exponentially localized NMs induce a dependence $\kappa_{linear} 
\sim {\rm e}^{-N/\xi}$. In the presence of nonlinearity the exponential decay with $N$ will stop at the
saturated value of $\kappa$. Assuming that $\kappa \sim T^{\alpha}$ it follows that the
system size for saturated conductivity values scales as $N \sim - \alpha \xi \ln T$.
Thus for lower temperatures the system size needed to reach saturated
values of $\kappa$ increases. This is due to the dropping of the saturated value of $\kappa$ with $T$.
We were able to 
reach temperatures as low as $T=0.005$ and conductivity values $\kappa = 0.008$.
According to Ref. \cite{lbksf10} (inset in Fig.3) the regime of strong chaos
is realized for $0.01 < T < 0.3$. Therefore we expect to observe $\kappa \sim T^2$
for $0.01 < T < 1$, followed by a maximum and a further dropping with increasing temperatures.
These predictions are indeed compatible with the numerical results in Fig.\ref{fig3}.
We also compute the conductivity for a reference ordered chain with $\tilde{\epsilon}_l=1$,
which coincides with the high temperature part of $\kappa(T)$ for the disordered case.
Therefore the observed maximum is due to disorder, as well as the decrease of $\kappa$ with further
lowering of the temperature. The low temperature data are also consistent with the $T^2$ law.
However the temperature window is too narrow to conclude about a confirmation. Also we expect
that at even lower temperatures (so far not reachable with our computational means) the conductivity
should cross over into the weak chaos regime and follow the $T^4$ law.

In order to find evidence for the asymptotic low-temperature behavior $\kappa \sim T^4$ we chose a stronger disorder
$W=6$. In that case $\xi \approx 3$ and $V\approx 10$. The conductivity is reduced by two orders of magnitude.
We are therefore quite limited in the temperature window, and do not obtain conclusive data (see Fig.\ref{fig3}).
We again find a maximum in the $\kappa (T)$ dependence, and a clear indication that the small temperature conductivity drops faster than
$T^2$ with temperature. Yet we do not have enough low temperature data to conclude about the exponent.

\subsection{FSW chain}

\begin{figure}[ht]
\begin{center}
\includegraphics[width=0.9\columnwidth]{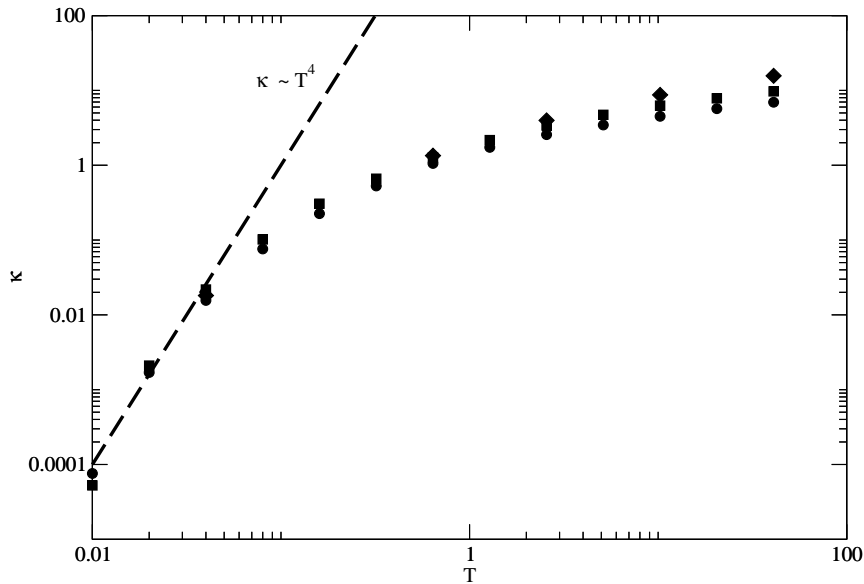}
\caption{FSW chain: $\kappa(T)$ for system sizes $N=16$ (filled circles), $N=64$ (filled
squares), and $N=256$ (filled diamonds). The dashed line corresponds to the law $T^4$.  }
\label{fig4}
\end{center}
\end{figure}

To observe the asymptotic low temperature behavior of the conductivity, we explore the strong disorder limit and study the FSW chain.
Since the localization length $\xi = 0$ for the FSW chain, we can use small system sizes
for small temperatures, while (as discussed above) we will need larger system sizes to
obtain the saturated conductivity values for larger temperatures. In Fig.\ref{fig4}
we plot the conductivity $\kappa$ versus temperature for different system sizes $N=16,64,256$. Indeed,
we observe a significant difference between the data for temperatures $T > 1$,
while the data agree well for lower temperatures.
Therefore we can reach lower values of the conductivity as compared to the KG chain case. 
Yet for $\kappa < 10^{-4}$
(which corresponds to $T \approx 0.01$) we again reach our computational limitations, since the number
of realizations and the integration times, which are needed in order to reduce the
impact of fluctuations, increase drastically.
The numerical results in Fig.\ref{fig4} indicate that for low temperatures the conductivity behaves as $\kappa \sim T^4$.
Also we observe at larger temperatures a crossover to the predicted pseudo-FPU regime.
Again we need more data in the low temperature regime to be more conclusive about the conductivity data.

\section{Discussion and outlook}

We computed the thermal conductivity for nonlinear disordered chains and obtained results which indicate that
theoretical predictions based on wave packet spreading are plausible. In particular we find indications for an asymptotic low-temperature
$\kappa \sim T^4$ and intermediate temperature $\kappa \sim T^2$ laws. To be more conclusive, we need data at even 
lower temperatures than those presented here. The needed CPU times are however currently exceeding our capabilities.
Possible solutions include the usage of massive parallel computing equipment, or the search for other models which
will allow to make the needed observations with standard computational equipment.
\\
Acknowledgments
\\
MI acknowledges the support of the Federal Program `Scientific and scientific-diductional brainpower of innovative Russia' for 2009-2013 (contracts $\Pi$2308, 02.740.11.0839)  and RFBR 10-02-00865.
We thank J. Bodyfelt, G. Gligoric, T. Lapteva, D. Krimer and H. Skokos for stimulating discussions.

\bibliographystyle{pramana}

\end{document}